\documentstyle[multicol,aps,epsf,rotate]{revtex}
\begin{document}
\input psfig.tex
\draft
\title{Self-Organized Criticality and Synchronization in the Forest-Fire Model}
\author{Barbara Drossel}
\address{Massachusetts Institute of Technology, \\ Physics Department Room
12-104, \\ Cambridge, MA 02139, USA}
\date{\today}
\maketitle
\begin{abstract}
Depending on the rule for tree growth, the forest-fire model shows either
self-organized criticality with rule-dependent exponents, or
synchronization, or an intermediate behavior. This is shown analytically
for the one-dimensional system, but holds evidently also in higher dimensions.
\end{abstract}

\begin{multicols}{2}
During the past years, systems which exhibit self-organized criticality (SOC)
have attracted much attention, since they might explain part of the abundance
of fractal structures in nature  \cite{btw}. Their common
features are slow driving or energy input (e.g. dropping of sand grains,
increase of strain, growing of trees) and rare dissipation events which are
instantaneous on the time scale of driving (e.g. avalanches \cite{btw},
earthquakes \cite{chr1}, fires \cite{dro0}).  In the stationary state,
the size distribution of dissipation events
obeys a power law, irrespective of initial conditions and without the need to
fine-tune parameters.

While the critical behavior of the sandpile model is relatively well
understood,
it is in general not clear, under which conditions SOC
behavior occurs. Models of the above-mentioned type show a variety of different
phenomena, depending on details of the model rules. When e.g. the form
of the driving force in the earthquake model \cite{chr1} is modified, a
transition from SOC to synchronization (activity bursts covering the whole
system) is found \cite{CPDA}. When in the same model the boundary conditions
are changed from
open to periodic, the SOC behavior is replaced by a periodic state
\cite{TangMiddleton}.
Another earthquake model, which includes inertia,
shows a superposition of a power law and a peak at large avalanches, where
most of the energy is released \cite{CarlsonLanger}.
A continuous forest-fire model \cite{BCJ}
shows periodic behavior or finite avalanches, depending on the value
of a parameter \cite{SGJ}.

Another question in the context of SOC deals with the universality of the
critical exponents. The critical behavior
of the sandpile model seems to be robust with respect to various changes,
and the critical exponents of the forest-fire model
do not depend on the lattice symmetry \cite{cla1,dro3} or the presence of a
further parameter \cite{dro4}. In contrast, the critical
exponents of the earthquake model depend continuously on the degree of energy
conservation \cite{chr1,JKG}. A similar dependence of critical
 exponents on the
model parameter is found in a new version of the forest-fire model with tree
conservation \cite{beyond}.

Since a great part of the above-mentioned work is performed numerically, there
is substantial need for analytically understanding these various phenomena.
In this paper, I study analytically a generalized version of the forest-fire
model in one dimension. Different rules for tree growth lead to either
SOC behavior with  rule-dependent exponents, or to synchronization, or to
an intermediate state with a superposition of a power law and a peak at large
fires. These results are confirmed by computer simulations.
Similar behavior is expected in higher dimensions.

The model is defined as follows: Each site in a one-dimensional system of
length $L$ is either occupied by a tree, or it is empty. Trees on neighboring
sites belong to the same tree cluster. During one time step of size $dt$, each
tree is struck by lightning with probability $f$. Trees struck by lightning and
all other trees in the same cluster burn down and turn to empty sites
immediately. Empty sites have a distribution of life times $P(\tau)$,
i.e. $P(\tau) dt $ is the probability that a site which just burned down will
remain empty exactly until the time $t$.  The
probability that a site which has been empty for a time $t$ becomes
occupied by a tree during the next time unit $dt$ is therefore
$ P(t)dt / \int_t^\infty P(\tau) d \tau$.
It is reasonable to require that the mean life time $\bar\tau$ of empty sites
is finite: $\bar \tau = \int_0^\infty P(\tau) \tau d\tau < \infty$. This means
that $P(\tau)$ decays faster than $\tau^{-2}$ for large $\tau$.
The condition that burning is
fast compared to tree growth places the model in the class of systems with slow
driving and instantaneous dissipation events.
The mean number of trees growing per unit time is $L(1 - \rho) / \bar \tau$,
and the mean number of lightning strokes per unit time is $L f \rho$, where
$\rho$ is the mean tree density in the stationary state.
The mean number of trees destroyed per lightning stroke consequently is
\begin{equation}
\bar s = (1 - \rho) / f \bar \tau \rho\, . \label{bars}
\end{equation}
When $f$ becomes very small,
the mean number of trees destroyed by a fire is large. If there are only large
fires but no small ones, large parts of the system burn down simultaneously and
may therefore be synchronized. If there are fires of all sizes up to a cutoff
size, the system is close to a critical point and shows scaling over many
orders of magnitude. In this paper, we always assume
that the system size $L$ is so large that no finite-size effects occur. The
cutoff is then a function of $f$ and $P(\tau)$ (see Eq.(\ref{nmax}) below).

 In the original forest-fire model, each empty site becomes occupied by a tree
 with probability $p$, and
 consequently $P(\tau) = p\,\exp(-p\tau)$. In the limit $f/p \to 0$, the size
 distribution of forest clusters in the stationary state is then essentially  a
 power law with a cutoff at sizes of the order $ (p/f) / \ln(p/f)$
 \cite{dro2}. In the following, we will
 determine analytically the size distribution of forest clusters for different
 $P(\tau)$.

To this purpose, let us consider a string of $k$ neighboring sites in the
system. If the lightning probability $f$ is sufficiently small, lightning does
not strike this string before all its trees are
grown. Starting with a completely empty state, the string passes therefore
through a cycle. Trees grow on the string, until it is
completely occupied by trees. Then the forest in the neighborhood of the string
will also be quite dense. The forest on the string is part of a forest cluster
which is much larger than $k$. Eventually that cluster becomes so large that it
is struck by lightning with a nonvanishing probability. Then the forest cluster
burns down, and the string again becomes completely empty, and the cycle
restarts. The mean time $T(k)$, which it takes for $k$ trees to grow on $k$
empty sites, satisfies
\begin{equation}
\int_{T(k)}^\infty P(\tau) d \tau \propto 1/k. \label{eqT}
\end{equation}
This equation can e.g. be obtained from the condition that the tree density is
larger than $1-1/k$ for $T=T(k)$. A more rigorous derivation gives the same
result.
We assumed that the probability that lightning strikes the string during that
time is negligeable. The maximum string size for which our considerations are
valid  is therefore  given by the condition
\begin{equation}
k_{\text{max}} f \int_0^{T(k_{\text{max}})} \left(T(k_{\text{max}}) -
\tau\right) P(\tau) d\tau  \ll 1. \label{nmax}
\end{equation}

After time $t$, an initially empty string of
$k<k_{\text{max}}$
sites is occupied by $m(t) = k \int_0^t P(\tau)d\tau$ trees on an average. For
not too small values of $m$, fluctuations around this mean value are relatively
small. The probability  ${\cal P}_k(m)$ to find $m$ trees on a string of
$k<k_{\text{max}}$  sites therefore satisfies for $1 << m < k$
\begin{equation}
{\cal P}_k(m) \propto \left[(d m / d t )_{t = t_{mk}}\right]^{-1} = \left(k
P(t_{mk})\right)^{-1}, \label{pnm}
\end{equation}
where $t_{mk}$ is the time after which the initially empty string is occupied
by $m$ trees.
Let $n(s)$ be the number of clusters of $s$ trees, divided by the  number of
sites $L$. $n(s)$ is identical to the probability that a string of $s+2$
neighboring sites is occupied by $s$ neighboring trees, with one empty site at
each end,
\begin{equation}
n(s) = P_{s+2}(s) / {s + 2 \choose s} \simeq 2 P_{s+2}(s) /s^2. \label{n(s)}
\end{equation}
For small $s$, there are deviations from this law due to the discreteness of
the lattice.
Remember that the result Eq.(\ref{n(s)}) is valid only for $s < s_{\text{max}}
= k_{\text{max}} - 2$. For larger $s$, clusters are struck by lightning with a
nonvanishing probability before they can grow to larger clusters, and therefore
Eq.(\ref{n(s)}) does not hold any more.

The mean number of trees destroyed per lightning stroke can be expressed in
terms of the cluster size distribution,
\begin{equation}
\bar s = \sum_1^\infty s^2n(s) / \rho = \sum_1^{s_{\text{max}}} s^2n(s) / \rho
+
\sum_{s_{\text{max}} + 1}^\infty s^2n(s) / \rho. \label{bars2}
\end{equation}
Here, we have to distinguish two different situations: Depending on the precise
form of $n(s)$, the second term can be neglected in the limit $f \to 0$ with
respect to the first term, or it dominates. In the first case, the system has
fires of any size and therefore is in a critical state, in the second case, the
dynamics are dominated by large fires. The first case will occur if $P(\tau)$
has a long-time tail. Then the neighborhood of the tree which is struck by
lightning is likely to contain surviving empty sites which eventually stop a
fire. The forest density approaches the value 1 in the limit $f \to 0$, since
otherwise $\bar s $ could not diverge. The second case will occur if tree
growth is finished after a finite time. Then large regions of the
system will become simultaneously occupied by a dense forest  and burn down
together. In this case the mean forest density will be smaller than 1.

\begin{figure}
\narrowtext
\centerline{\rotate[r]{\epsfysize=3in
\epsffile{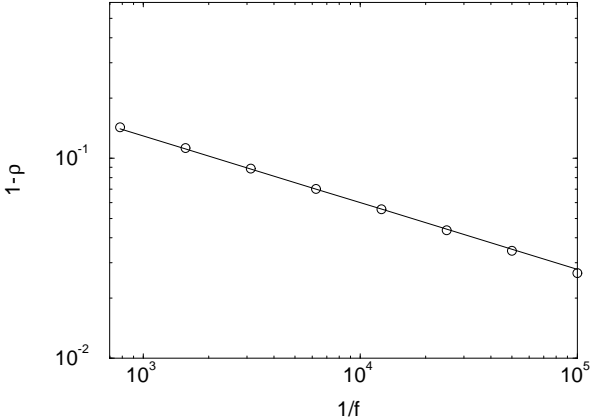}}}
\vskip 0.5true cm
\caption{Density of empty sites as function of $1/f$ for $L = 10000$ and
$\alpha = 3$. The solid line has the slope $-1/3$.
}
\label{dichte}
\end{figure}

To make these statements more precise, let us specify the different possible
cases:

(i) $P(\tau)$ has a long-time tail, i.e.
$ P(\tau) \propto \tau^{-\alpha} $
for large $\tau$, with $\alpha > 2$. Sites which have been empty for a time
$t$ become occupied by a tree during the next time step with probability
$(\alpha - 1) dt / t$, where
$\alpha$ is now a parameter and not an exponent. From Eqs.(\ref{eqT}) and
(\ref{nmax}) we
find $T(k) \propto k^{1/(\alpha - 1)}$ and $s_{\text{max}} \propto f^{-1 +
1/\alpha}$. With Eqs.(\ref{pnm}) and (\ref{n(s)}) we obtain the size
distribution of forest clusters
\begin{displaymath}
n(s) \propto s ^{-2 + 1 / (\alpha - 1)}.
\end{displaymath}
The forest density $\rho = \sum_{s = 1}^\infty sn(s)$ cannot be larger than 1,
and consequently $n(s)$ assumes a normalization factor
$s_{\text{max}}^{-1/(\alpha - 1)} \propto f^{1/\alpha}$.
\begin{figure}
\narrowtext
\centerline{\rotate[r]{\epsfysize=3in
\epsffile{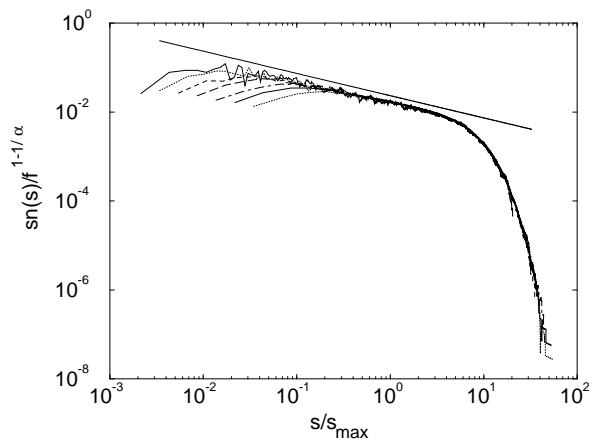}}}
\vskip 0.5true cm
\caption{Scaled size distribution $sn(s)/f ^ {1 - 1 / \alpha}$ of fires as
function of $s / s_{\text{max}}$. The parameters are $L = 10000$ and $\alpha =
3$, for $1/f = 100000$ (solid), 50000 (dotted), 25000 (dashed), 12500 (long
dashed), 6250 (dot-dashed), 3125 (solid), 1562.5 (dotted). The straight line
has the slope $-0.5$.
}
\label{tail}
\end{figure}
With Eq.(\ref{bars2}) we find $\bar s  \propto s_{\text{max}}$, which leads
with Eq.(\ref{bars}) to $1 - \rho \propto f ^ {1 / \alpha}$. These results
represent SOC behavior, and the values of the exponents depend on $\alpha$,
i.e. they are non universal. The analytic results are confirmed by
simulations. Fig.\ref{tail}  shows the collapsed curves for the size
distribution $sn(s)$ of fires for $L = 10000$ and  $\alpha = 3$, for different
values of $f$, confirming the relations for $n(s)$, for $s_{\text{max}}$, and
the normalization factor $ \propto f^{1/\alpha}$.  Fig.\ref{dichte} shows the
density of empty sites as function of $f$, confirming the relation $1 - \rho
\propto f ^ {1 / \alpha}$.

(ii)  $P(\tau)$ decays exponentially fast: $P(\tau) \simeq \exp(-\tau^\beta)$
for large $\tau$. This situation comprises also the original forest-fire model.
The exponents are obtained from case (i) by taking the limit $\alpha \to
\infty$. In addition, there occur logarithmic corrections. With
Eqs.(\ref{eqT}), (\ref{nmax}), and (\ref{n(s)}), we find $T(k) \simeq
(\ln(k))^{1/\beta}$, $s_{\text{max}} (\ln(s_{\text{max}}))^{1/\beta} \propto
1/f$, and $n(s) \propto s^{-2}/\ln(1/f)$. For $\beta = 1$, these results have
already been analytically derived and confirmed by simulations in  \cite{dro2}.

\begin{figure}
\narrowtext
\centerline{\rotate[r]{\epsfysize=3in
\epsffile{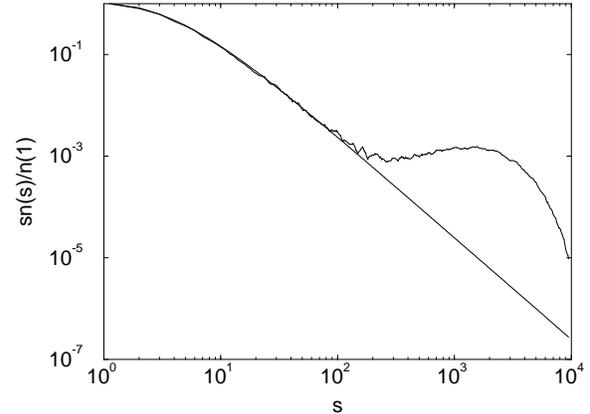}}}
\vskip 0.5true cm
\caption{Normalized size distribution of fires for $P(\tau) = 1/100$,  $T_0 =
100$, $L = 10000$, $f = 0.00001$. The smooth line is the theoretical result
Eq.(5), which is valid for $s < s_{\text{max}}$.
}
\label{T}
\end{figure}

(iii) All trees grow within a time $T_0$, i.e. $P(\tau > T_0) = 0$ and
 $\int_0^{T_0} P(\tau) d \tau = 1$.
In this case $k_{\text{max}}$ is essentially given by $k_{\text{max}} \propto
1/fT_0$. The large-scale dynamics of this system can be obtained by performing
a scale transformation: $k_{\text{max}}$ sites form together a big site which
becomes occupied exactly after $T_0$ steps, and the lightning probability for
this coarse-grained system is $k_{\text{max}} f \propto 1 / T_0$. This is a
model with deterministic tree growth, which shows synchronization and will be
considered in the next paragraph. Before, let us take a look at the short-scale
dynamics: The size distribution of clusters smaller than $s_{\text{max}}$ is
obtained in the same way as before. If $\lim_{\tau \to T_0} P(\tau) \neq 0,
\infty$,  we find $n(s) \propto s^{-3}$ and consequently
$\sum_1^{s_{\text{max}}} s^2n(s) / \rho \propto \ln(1/fT_0)$. Since we know
(Eq.(\ref{bars})) that $\bar s \propto 1/fT_0$, we conclude from
Eq.(\ref{bars2}) that most fires are large and are not described by the power
law $sn(s) \propto s^{-2}$. Fig.\ref{T} shows
 the size distribution of fires for $P(\tau) = 1/T_0$,  $T_0 = 100$, $L =
10000$, and $f = 0.00001$. In addition to the mentioned power law, there is a
peak at large fires, indicating that large segments of the system burn down
together.

The special case $\lim_{\tau \to 0}P(T_0 - \tau) \propto \tau^\alpha$ for
$\alpha > 0$ or $-1 < \alpha < 0$ leads to $n(s) \propto s^{-3 + \alpha / (1 +
\alpha)}$.

\begin{figure}
\narrowtext
\centerline{\rotate[r]{\epsfysize=3in
\epsffile{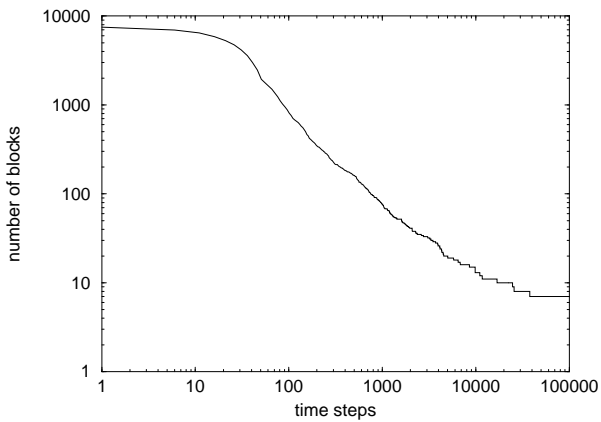}}}
\vskip 0.5true cm
\caption{Number of synchronized blocks as function of time for $L = 10000$, $f
= 0.005$, and $T_0 = 50$.
}
\label{syn}
\end{figure}

(iv) Last, let us consider the case of deterministic tree growth $P(\tau) =
\delta(T_0 - \tau)$, which is also the coarse-grained version of situation
(iii). Here it is easy to understand how synchronization arises: If two
neighboring sites happen to be both occupied by a tree, they  burn down during
the same fire. Consequently, both trees  regrow simultaneously and burn down
simultaneously for all future times. So the system consists essentially of
synchronized blocks which turn to trees  and burn down
simultaneously. Neighboring blocks join, if the block which grows first is not
struck by lightning before its neighbor grows. So the number of blocks
decreases with time, and their mean size increases. In the stationary state,
the mean block size is so large that each block is struck by lightning
immediately after it grows, so that neighboring blocks cannot join any
more. This stationary state is periodic with a period $T_0$. In the limit $f
T_0 \to 0$, the block size becomes very large, and eventually the whole
system is synchronized and fires every $T_0$ time steps. Fig.\ref{syn} shows
the number of blocks as function of time for $L = 10000$, $f = 0.005$, and $T_0
= 50$, starting with a random state.

Generalization to higher dimensions is straightforward: It is clear that the
case $P(\tau) = \delta(T_0 -\tau)$ leads to synchronization of large blocks of
the system in any dimension. On the other hand, it is known that the system is
SOC for $P(\tau) = p\,\exp(-pt)$. Therefore, it can be expected that when
varying $P(\tau)$ the critical exponents change their values, and that finally
the system becomes synchronized. However, the calculations performed in this
paper cannot be applied to higher dimensions. In contrast one dimension, fire
can now burn regions which are not completely dense, and part of the
trees in that region survive the fire, thus generating correlations which are
not present in the one-dimensional model.

To summarize, I have shown that depending on the rule for tree growth the
stationary state of the forest-fire model shows either SOC or synchronization,
or a superposition of both. In the SOC
state, the critical exponents vary continuously when $P(\tau)$ is
changed. These results are similar to those for the earthquake model which
shows SOC or synchronization depending on the form of the driving force
\cite{CPDA}.
In \cite{TangMiddleton}, it is shown that the SOC behavior in the
earthquake model is related to the tendency of neighboring sites to
synchronize. Neighboring sites which topple during the same avalanche are
likely to topple also the next time during the same avalanche. The forest-fire
model shows a similar feature in the SOC state: A region which has just been
crossed by a fire is empty (in one dimension) or almost empty (in two
dimensions). This
region is with a high probability again covered by an almost
dense forest in the moment when it is ignited. So neighboring sites which
burned down
during the same fire are
likely to burn down simultaneously also the next time. Due to the
randomness of tree growth, this ``synchronization'' is not perfect. The degree
of synchronization increases when tree growth becomes more deterministic, first
showing a peak at large fires, and finally perfect synchronization of large
clusters.

When the stochastic rule for lightning strokes is replaced by a more
deterministic one, the model also becomes more synchronized. Lightning might
strike only trees which
have a certain age, or only clusters which have a certain size. In these cases,
there are only large fires in the stationary state and no small ones.
However, the size distribution of forest clusters smaller than a cutoff size
is not considerably affected by these changes since Eq.(\ref{pnm}) was derived
under the only assumption that lightning does not strike a string of size $k <
k_{\text{max}}$, before all its trees are grown.

The results of this paper might have implications for excitable media
\cite{tys} (e.g. spreading of diseases,
autocatalytic chemical reactions, propagation of electrical activity in neurons
or heart muscles). These systems essentially have three states which are called
quiescent, excited, and refractory. If there is an excitation, it  spreads to
the quiescent neighbors. After excitation, a region is refractory to further
stimulation and needs some time to recover its quiescent state. The model
discussed in this paper describes an excitable medium in  the limit where
excitation spreads fast on the scale of the refractory time, and where
spontaneous excitation is rare. In cases where this limit can be realized
experimentally, such a system should show either synchronization or SOC,
depending on whether the transition from the refractory to the quiescent state
and from the quiescent to the excited state are deterministic or stochastic.

I thank S. Clar and one of the referees for their comments on the manuscript.
This work was supported by the Deutsche
Forschungsgemeinschaft (DFG) under Contract No. Dr 300/1-1, and by the
NSF grant No. DMR-93-03667.

\end{multicols}

\end{document}